# Characterization of Flux Trapping in and Fabrication of Large-Scale Superconductor Circuits Using AC-Biased Shift Registers With 108500 Josephson Junctions

Evan B. Golden, Neel A. Parmar, Vasili K. Semenov, and Sergey K. Tolpygo, *Senior Member, IEEE*

*(Invited Paper)*

*Abstract*— A variety of superconductor integrated circuits comprising six ac-powered SFQ shift registers with a total of 27078 bits and 108500 Josephson junctions (JJs) per 5 mm x 5 mm chip have been designed, fabricated, and tested to characterize flux trapping, fabrication process yield, and parameter spread. The six 4513-bit registers in the circuits have a common single-phase ac clock and individual input/output drivers enabling their parallel testing. We have investigated flux trapping in the circuits with various geometry, size, and distance between moats in two active ground planes (GPs), and containing up to three additional 'dummy' GPs, using multiple cooldowns through the critical temperature with various cooling rates and residual magnetic fields up to ≈1.2 µT. For the slit-type and square moats arrayed along the sides of the register cells, we have found a negligible effect of flux sequestered in the moats on the operating margins of the registers, and negligible probability of detrimental flux trapping outside of the moats. Circuits with 0.3-µm-wide slit moats occupying <2% of the circuit area were fully operational in 100% of cooldowns, supporting the viability of VLSI superconductor digital circuits. We have found a strong enhancement of flux trapping outside of the moats in circuits with closely spaced GPs and determined a critical distance, $t_c$=0.6 µm, between them. The presence of GPs spaced below $t_c$ rendered the circuits nonoperational in 100% of cooldowns. We have measured 30 chips with >3M JJs and determined individual cell margins in 138 registers to characterize the fabrication-related parameter spread and detect fabrication defects and flux-trapping events. By finding outlier cells in the statistical distribution of the individual cell margins, we detected about one defect per million JJs, in most cases causing magnetic flux trapping in the affected cell.

The circuits have been fabricated in the SFQ5ee fabrication process at MIT Lincoln Laboratory (MIT LL).

## I. INTRODUCTION

WHILE superconductor electronics has been developed for many decades, the technology has yet to gain a foothold in high-performance computing. The complexity, and subsequently functionality and utility, of superconductor circuits is typically constrained by the achieved integration scale - the number of Josephson junctions (JJs) contained within a circuit. Achieving beyond-CMOS computing abilities with superconductor electronics will require circuits containing millions of Josephson junctions and other circuit components. There have been very few demonstrations of circuits with this complexity, mainly quantum annealing processors from D-Wave Systems, Inc. [1], and single flux quantum (SFQ) shift registers [2], whereas the typical scale of the demonstrated circuits does not exceed a hundred thousand JJs [3], [4], [5], [6], [7], [8], [9] and references therein. This limitation is imposed by a variety of factors including the ability to design and fabricate large-scale circuits, challenges with their powering and clock distribution, and magnetic flux trapping [10], [11], [12].

Shift registers in a variety of superconductor logic families (RSFQ, RQL, AQFP) were used as a diagnostic circuit for characterizing the integrated circuit fabrication process yield, parameter spreads, and propensity to flux trapping [2], [4], [6], [7], [13], [14], [15], [16]. In our previous publications [2], [5], we developed a scalable to millions of JJs SFQ shift register using a single-phase ac clock power, the so-called ac-powered or ac-biased shift register. We demonstrated how operating margins of the individual bits (cells) of this very long shift register can be extracted and used to characterize the fabrication process parameter spreads and detect flux trapping inside the cells and in the ground plane moats guarding the cells.

In this work, we modified the design of the ac-powered shift register by splitting it into six shorter registers with independent input/output circuitry, allowing for much faster measurements of the operating margins of the individual cells. We measured over 30 chips using multiple thermal cycling and different cooling rates to investigate flux trapping probability and efficiency of various moat shapes and sizes in protecting the circuits against detrimental flux trapping inside the bit cells rather than in the guarding moats. These measurements were also used to study the effects of the number of ground planes, their width, and the moats' noncongruence on flux trapping, detect fabrication defects, and characterize the fabrication process yield and parameter spreads. By studying 180 registers with over three million JJs, we detected about one fabrication-related defect per million JJs, typically causing flux trapping in the affected cell, but not necessarily rendering the whole circuit nonoperational.

In Sec. II we give a background on flux trapping, scaling of superconductor electronics, and the use of shift registers. In Sec. III we present the design and salient features of the ac-powered circuits used in this research, describe the test procedure and the cell margin extraction method. The obtained results are given in Sec. IV for the registers with different sizes and shapes of the moats, and different numbers of the ground planes layers utilized in the circuit designs. This is followed by the Discussion and Conclusion.

## II. BACKGROUND AND PRIOR WORK

*A. Flux Trapping*

Very large-scale integration (VLSI) of superconductor circuits requires fabrication with minimal variation of critical currents of JJs and linewidths of inductors in order to provide an acceptable circuit yield. To demonstrate the capabilities of a fabrication process, individual circuit elements, such as Josephson junctions, resistors, inductors, vias, etc., can be fabricated and characterized, but their statistics might not capture interactions that occur in VLSI circuits. These interactions include: changes in metal density and proximity effects impacting linewidth and dielectric thickness, and thus inductance; local stresses and hydrogen contamination affecting junctions' critical currents; galvanic corrosion and related fabrication defects occurring under certain circuit layout configurations; particles, etc. These are hard defects that become imprinted into the circuits after their fabrication.

Specific only to superconductor electronics, there is another type of defects – magnetic flux trapped inside the superconducting films in the form of Abrikosov vortices and inside multiply connected superconducting areas and inductor loops in and between the circuit layers in the form of quantized fluxons.

Vortices form in superconducting films, which are type II superconductors [17], during a circuit cooldown in a nonzero residual magnetic field, $H_r$, below a temperature $T_{c2} \cong T_c$ at which $B_{c2}(T_{c2}) = \mu_0 H_r$, where $B_{c2}(T) = \Phi_0/(2\pi\xi(T)^2)$ is the second critical magnetic field of the superconducting film and $\xi(T)$ is the temperature-dependent coherence length [18]. Below a certain temperature, $T_{c1}$, also very close to the $T_c$, the existence of vortices in the films becomes energetically unfavorable when $\mu_0 H_r$ becomes smaller than the first critical magnetic field, $B_{c1}(T)$.

In bulk superconductors

$$B_{c1}(T) = \frac{\Phi_0}{4\pi\lambda(T)^2}\left(\ln\frac{\lambda(T)}{\xi(T)} + 0.08\right), \tag{1}$$

where $\lambda(T)$ is magnetic field penetration depth [17], [18], [19], [20]. In thin films, $B_{c1}(T)$ depends on the film shape, thickness, and other parameters [21], [22], [23], [24]. As a result, analytical expressions were obtained only for a few cases - an infinitely long narrow strip [22], [24], a disk [23], or a ring [26]. E.g., for narrow strip of width $W \ll \Lambda$ in perpendicular magnetic field [22]

$$B_{c1}(T) = \frac{2\Phi_0}{\pi W^2}\ln\frac{2W}{\pi\xi(T)}, \tag{2}$$

where $\Lambda = 2\lambda(T)^2/d$ is the Pearl length [21]. Numerical modeling is required in all practical cases of patterned films in integrated circuits [12].

In addition, there is usually a surface barrier which prevents expulsion of vortices from superconductors due to attraction of vortices to superconductor boundaries and surfaces, e.g., edges of the film, as first shown by Bean and Levingston [27]. There are other details related to the edge barrier and to whether vortices at $T_{c2}$ are formed as single vortices or as vortex-antivortex pairs [28], [29], which affects their expulsion at a lower temperature and changes the value of critical magnetic field for narrow strips to either [22], [30]

$$B_0 = \frac{\pi\Phi_0}{4W^2}, \tag{3}$$

or [31], [32]

$$B_K = \frac{1.65\Phi_0}{W^2}. \tag{4}$$

Both of these expressions are temperature- and material-independent as a result of the London approximation used to derive them [33]. All these factors make finding complete flux expulsion conditions, e.g., expulsion field $B_{exp}$ for vortices, or vortex-antivortex pairs, from films with complex shapes and patterns used in integrated circuits difficult and often ambiguous.

Yet another very important, and often dominant, factor is vortex pinning by defects, grain boundaries, and film nonuniformities, strongly preventing their expulsion. Vortex pinning strength in the circuit films depends on the material's properties and microstructure, e.g., nonuniformity of the $T_c$ and the mean free path [33], and of the penetration depth [34], and can change depending on the fabrication conditions. As a result of many factors, partially listed above, vortices may become trapped inside the films in metastable states despite that $\mu_0 H_r < B_{exp}$. The location and amount of the trapped flux depend on $H_r$, cooling rate through $T_c$, films' microstructure and patterns, nonuniformity of the films' properties, and may change from cooldown to cooldown.

*B. Flux Trapping Protection: Moats and Moat Shapes*

A standard approach to protecting circuits against flux trapping inside the logic cells and on or near the JJs has been since 1983 the use of guarding moats in the ground planes of the circuits [34], [35], [36], [37], [12]. Moats within superconductor circuits are areas where the superconductor film has been selectively removed to sequester magnetic flux that could otherwise be trapped in the film during cooling through the critical temperature of the superconductor. The film and moat boundaries attract Abrikosov vortices forming inside the circuit films upon field cooling through $T_c$. The vortices are mobile near $T_c$ and can move towards moats either viscously or by thermally assisted hopping between pinning centers in the film. Then, the flux becomes sequestered in moats far away from any circuit parts which could be affected by the associated magnetic field and circulating currents; see [12] and references therein. This is a *good, intended, flux trapping* that needs to be distinguished from the unintentional and detrimental flux trapping inside the circuitry.

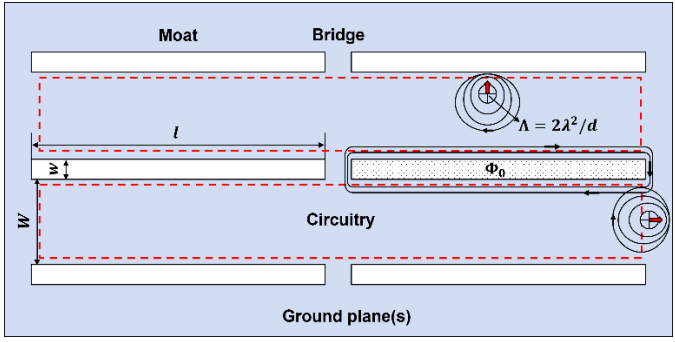

Fig. 1. A sketch of the top view of long slit-like moats used in [2], [5], [12], [40]. The moats are congruent in electrically connected by superconducting vias the bottom and top ground planes of a circuit. The moat length is $l$, pitch $p$, and width $w$. The distance between the parallel moats $W$ is set by the height (the $y$-direction size) of the circuit cells. The bridge length in the $x$-direction, $p - l$, is between 1 and 4 μm. Routing between the cells in the y-direction is typically done over the bridges and sometimes can be done over the moats if a dramatic increase in inductance of this path [45], [46] is acceptable. A moat with trapped flux quantum is shown. Abrikosov (Pearl) vortices attract to the empty moats and edges of the superconducting film but repel from the flux of the same polarity trapped in the moats.

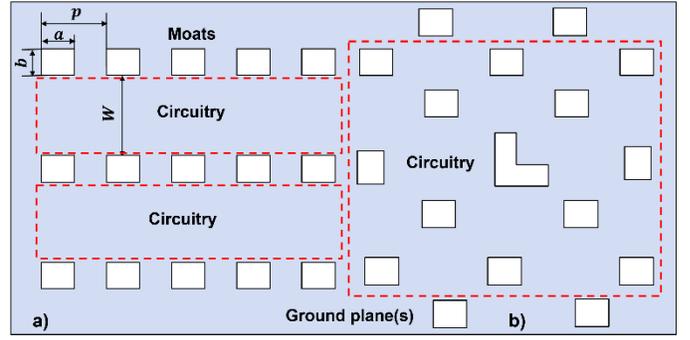

Fig. 2. A sketch of the top view of different design styles of rectangular or square moats: a) a regular pattern of moats outside the circuitry shown by a dash rectangle used in some circuit in this work and in [40]; b) a mixed pattern combining ordered and more or less randomly placed moats of various shapes occupying available empty space within and around the circuitry; used, e.g., in [13], [14], [15], [39], [40], [42], [43].

There has been a limited amount of research devoted to studying the effectiveness of different moat shapes and dimensions in flux sequestration and to the probability of flux trapping in various circuits [12], [13], [14], [39], [40], [41]. Some researchers [2], [5], [12], [39], [41] use slit-like moats splitting the wide ground plane films into long and narrow superconducting strips connected by bridges as shown in Fig. 1. This moating creates multiple long edges of the films for attracting vortices. Other design teams use square and rectangular moats or more complex shapes [6], [13], [14], [15], [16], [38], [40], [41], [42], [43]; see Fig. 2.

Slit-like moats have a large inductance, $L_{moat} \sim \mu_0 l$, proportional to the moat length, $l$. As a results, the total current circulating around the moat with a trapped flux quantum $\Phi_0$, $I_{cir} = \Phi_0/L_{moat} \propto l^{-1}$, becomes much smaller than the typical current of Josephson junctions used in the circuits, $I_c \gtrsim 30$ μA at $l \gtrsim 40$ μm. For comparison, square moats with side $a = 5$ μm, often used in superconductor circuit design, have $L_{moat} = 7.86$ pH and $I_{cir} \approx 0.26$ mA $\gg I_c$.

When moats fail to sequester all vortices and some of them remain trapped inside the circuitry, a *bad flux trapping occurs*.

### B. Moat Number Density and Distance Between Moats

While a vortex attracts to the moat edge, it repels from the flux of the same polarity trapped in the moat, reducing the attraction force and making flux-containing (filled) moats less efficient. The moat filling with flux may continue until the Gibbs free energy of the film with a vortex is larger than the free energy of the film without the vortex but with the extra flux quantum inside the moat. These free energy calculations are very complicated for arbitrary film and moat geometries, and were done only in a few cases [12]. Hence, developing simple intuitive moat design recommendations would be useful. For instance, the free energy requirement is always fulfilled if the total number of moats, or moat number density, $n_{moat}$, is equal or larger than the total number of vortices, or the vortex density $n_v$, that can exist in the ground planes of a circuit cooled in a given residual magnetic field, $B_r$,

$$n_{moat} \geq n_v(B_r) \qquad (5)$$

The latter can be estimated as $n_v \approx B_r/\Phi_0$. For the typical values $B_r \equiv \mu_0 H_r$=100 nT (1 mG) and 1 μT in, respectively, a three-layer mu-metal shield [44] and a single-layer mu-metal shield used for testing superconductor integrated circuits, $n_v \approx 5 \cdot 10^3$ cm$^{-2}$ and $5 \cdot 10^4$ cm$^{-2}$, respectively.

Although (5) overestimates the minimum number of the required moats, it not sufficient. No more than one of the vortices located along the moat side can be sequestered because the vortex-moat interaction is relatively short-ranged and attraction force to the next moat is usually small. Hence, the number of vortices on the typical area guarded by a moat (between two adjacent moats) should be

$$n_v W l \leq 1 \text{ or } B_r W l / \Phi_0 \leq 1, \qquad (6)$$

where $W$ is the distance between two parallel slit-like moats, i.e., the effective width of the ground plane between the moats; see Fig. 1. For $B_r$=10 μT and $l = 40$ μm, (6) requires $W \leq 50$ μm.

Well below the critical temperature, a vortex located a distance $a \gg \lambda$ away from a moat or the strip edges has an exponentially weak attraction to them, $\propto \exp(-2 a/\lambda)$ because the screening current decay length in films with thickness $d \gg \lambda$ is the magnetic field penetration depth $\lambda$. The penetration depth in Nb films at 4.2 K is $\lambda \approx 90$ nm [45], [46], and the film thickness in the SFQ5ee fabrication process used in this work is $d$ =200 nm [48]. However, very close to $T_c$ where $\lambda(T) \gg d$, an Abrikosov vortex becomes a Pearl vortex [21] with slowly decaying as $r^{-1}$ screening currents and attraction force to the film edge $f \sim \mu_0 \Phi_0^2/(4a^2)$ [21], [24] at $r, a \gg \Lambda$, where $r$ is the distance from the vortex center and $\Lambda = 2\lambda(T)^2/d$.

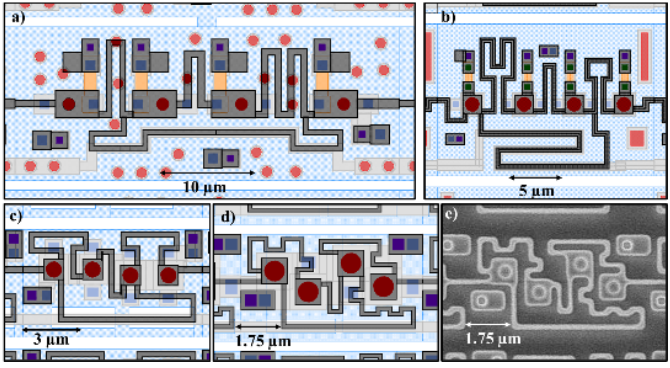

Fig. 3. Layouts of four versions of the ac-powered shift register unit cell fabricated previously at MIT LL [51], to demonstrate the cell area reduction with reducing the minimum linewidth and JJ area: (a) the cell [5] fabricated in the standard SFQ5ee fabrication process [48] with Josephson critical current density $J_c$ of 100 µA/µm$^2$ and 700 nm minimum linewidth; (b) the cell [2] fabricated in the SFQ5ee fabrication process with $J_c$ of 100 µA/µm$^2$ and 400 nm linewidth; (c) the cell fabricated using 250 nm minimum linewidth in the SC1 process [49], [50], the cell dimensions are 12 µm × 8 µm; (d) the FLXSHTL60 cell fabricated using 250 nm minimum linewidth and self-shunted JJs with $J_c$=600 µA/µm$^2$, the cell size is 9 µm × 6 µm [50]; (e) the most advanced 150nmv1 cell fabricated using 150 nm minimum linewidth and $J_c$ =600 µA/µm$^2$, the cell dimensions are 7 µm × 4 µm [50].

Using the standard two-fluid expression $\lambda(T) = \lambda(0)/(1 - T^4/T_c^4)^{1/2}$, $\lambda(0)$=81 nm [47], and $T_c = 9.1$ K, the temperature region where the Pearl-type vortices may exist in Nb films with $d = 200$ nm is very narrow, $1 > T/T_c > 0.956$, i.e., $T_c - T <$ 0.4 K . Cooling down through this region should be done slowly to allow vortices migrate to the moats or edges of the film, driven by the weak attraction force.

On the other hand, for macroscopic distances between the vortex and the film (or moat) boundary $a \sim W/2$, the attraction force is significant only if $a < \Lambda$, i.e.,

$$2\lambda(T)^2/d \gtrsim W/2 . \quad (7)$$

This takes place in the temperature interval

$$1 > (T/T_c)^4 \gtrsim 1 - \frac{4\lambda(0)^2}{dW} . \quad (8)$$

For the estimated above $W = 50$ µm, this interval is extremely narrow $T/T_c > 0.9993$ or $T_c - T <$0.006 K, that is much smaller than the typical superconducting transition width of Nb films caused by the film's nonuniformities. Hence, using this distance between the moats is impractical because vortices located far from the moats cannot be attracted to and sequestered in them. A more practical would be $W$ =15 µm or 20 µm. Indeed, at $W$ =15 µm, the temperature range of a long-range vortex-moat interaction is $T/T_c > 0.9978$ or $T_c - T <$ 0.02 K that is wider than the typical transition width. Cooling down through this temperature interval should be done slowly to allow vortices from the film interior migrate to the moats or edges of the film.

In addition to the diminishing with decreasing temperature attraction force, there is a competing vortex pinning force which increases with lowering temperature. As a result, at some temperature $T_f$, very close to $T_c$ and dependent on the film's microstructure, i.e., on the vortex pinning strength, the pining force becomes larger than the attraction force; vortices become pinned by defects and can only hop between pinning sites by thermal activation. Such a situation results in the "bad flux trapping" in the interior of the film.

The impact of the trapped flux depends on its location and the circuit design. It can be extremely detrimental, often rendering circuits completely nonoperational or requiring multiple cooldowns to expel the flux. The relevant films' parameters, $H_r$, and the number of cooldowns required to get a particular circuit operational are practically never reported in the publications. With increasing the scale of integration, all circuit components move closer to each other leaving less and less free space for flux sequestering moats and increasing moat-to-device coupling; the problem of bad flux trapping becomes more severe.

*C. Shift Registers as Flux Trapping and Fabrication Yield Diagnostics*

An ideal circuit for wholistically characterizing fabrication yield, parameter spreads, and flux trapping should be able to scale to VLSI, provide predictable behavior to discern correct operation or identify errors, and be simple to test. The ac-powered SFQ shift register has been previously demonstrated as a good diagnostic circuit due to numerous beneficial qualities: it utilizes all standard components available in the fabrication process; it can be easily modified by changing a single, simple, unit cell; it is powered by an inductively-coupled single-phase ac clock signal, alleviating any issues caused by dc biasing of large circuits. This enables scaling the circuit to arbitrarily large sizes with relative ease, limited only by the available chip size.

Multiple versions of ac-biased shift registers were designed [51], fabricated in various fabrication processes at MIT LL, and measured; see Fig. 3. The standard SFQ5ee process [48] has 500 nm minimum linewidth, which was reduced to 250 nm [49] and to 150 nm linewidth in the SC2 process using 193-nm photolithography [50]. Externally shunted Josephson junctions with a critical current density, $J_c$ of 100 µA/µm$^2$ in the SFQ5ee can also be substituted with self-shunted JJs with $J_c$ =600 µA/µm$^2$ to eliminate shunt resistors and further reduce circuit cell size [47], [50].

### III. CIRCUIT DESIGN, FABRICATION, AND TESTING

*A. Base Circuit Design: FLXTRP1*

Of all available options for the benchmark circuit design shown in Fig. 3, we selected a more aggressive version of the well-proven SFQ5ee process, using inductor linewidth of 400 nm and 500 nm square interlayer vias. The cross section of the process is shown in Fig. 4.

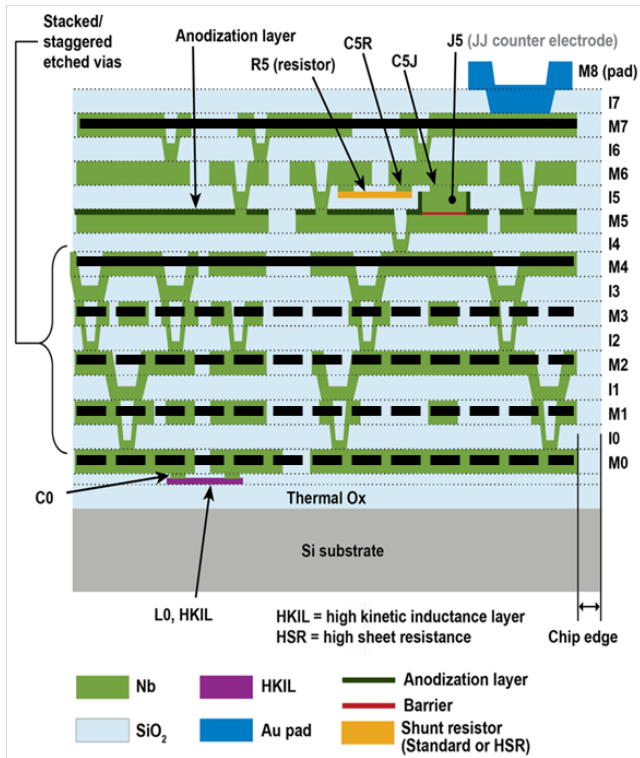

**Fig. 4**. Schematic cross section of the SFQ5ee fabrication process with eight Nb planarized layers [48]. Layers M4 and M7 were used as the bottom and top ground planes (marked by the black lines), respectively. They are connected by superconducting vias I4, I5, and I6. All Josephson junctions and inductors of the shift registers are sandwiched between these two ground planes for a better shielding. Various patterned shapes in the bottom Nb layers, M0-M3 were used as a metal fill for the dielectric planarization steps and as 'dummy' ground planes to investigate their effect on flux trapping in the circuits.

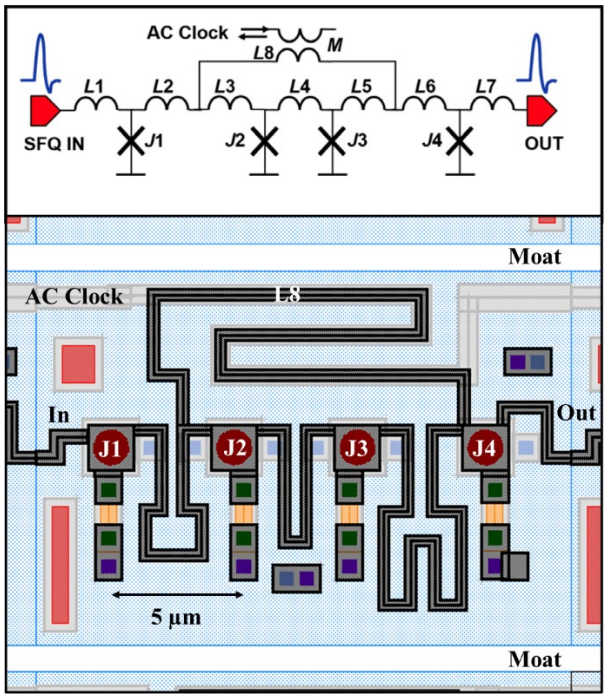

**Fig. 5**. The circuit diagram (top panel) and the layout (bottom panel) of a unit cell of the ac-powered shift registers [2], [5] used in this work. Long slit-type moats are clearly visible. The moats are congruent in the M4 and M7 ground planes of the circuit.

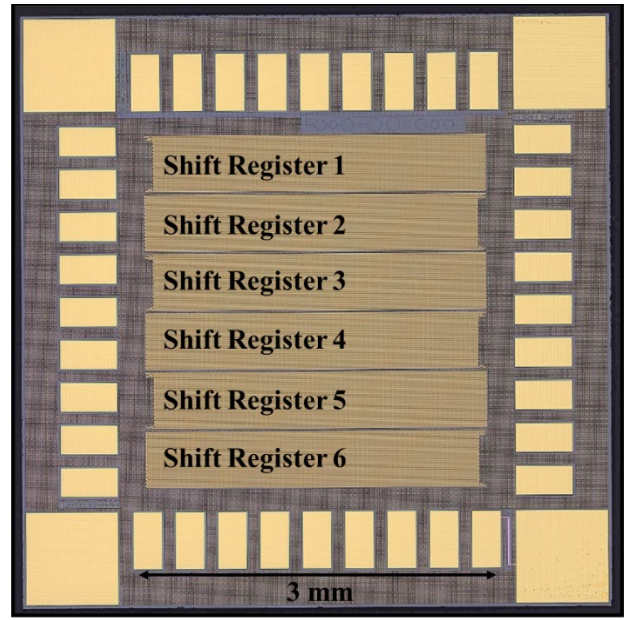

**Fig. 6**. Top view of the fabricated 5 mm x 5 mm chip with the diagnostic circuit FLXTRP1 consisting of six ac-powered shift registers and a 10-loop SQIF magnetometer, above the Register 1, for measuring the local magnetic field after the chip cooldown.

We used an approach described in [2]. To reduce the testing time and be able to test circuits with significant flux trapping or even partially defective, we changed the global design of the ac-powered shift register in [2]. Instead of using a very long shift register with a single input and multiple intermediate output taps, we split it into six independent shift registers each having its own data input and output. All six registers have a common ac-power (clock) delivery stripline and independent biases for the input/output (I/O) drivers. These features allowed us to increase the testing speed by doing a parallel testing of the six registers, and even test chips with fabrication defects, that would be impossible with a long sequential register where a single defect may completely block the data propagation.

The unit cell of the shift register contains four junctions; its electrical schematic, identical to the one used in [2], and layout are shown in Fig. 5. The cell dimensions are 20 μm × 14 μm. Each of the six individual shift registers is organized in 32 rows of 140 bit-cells connected in a serpentine manner by 31 U-turn bit-cells with identical schematics but different layout. The spacing between the rows was 1 μm (placement pitch for rows of 15 μm) to accommodate the slit-type moats shown in Fig. 1 with $w = 1$ μm between the rows. This resulted in the size of the U-turn cells of 19 μm x 29 μm. The first and the last rows in the register have an extra cell connected, respectively, to the Input DC-SFQ and the Output SFQ-DC converters. These two cells experience an additional dc bias current from the adjacent I/O drivers, making them special cells from a statistical point of view. Each of the six registers has an intermediate data output tap after the first 281 cells, using a U-turn bit-splitter cell #282, for fast screening and troubleshooting. The U-turn and the splitter cells have different designs from the row cell in Fig. 5b due to the physical constrains on the cells, and consequently have somewhat different operating margins and different flux trapping propensity. These make 31+3=34 special cells in the register having 4513 bits in total and 18,052 JJs. The circuit, referred to as "FLXTRP1," was designed on a 5 mm x 5 mm chip; it contains about 108,500 JJs in the entire circuit, including I/Os, with $1.33 \cdot 10^6$ JJ/cm$^2$ device density.

TABLE I
MOAT PARAMETERS OF THE BASE CIRCUIT FLXTRP1

| Circuit FLXTRP1 | Moat length, $l$ (µm) | Moat pitch, $p_x$ in x-direction (µm) | Bridge length between moats (µm) | Moat width, $w$ (µm) | Moat pitch, $p_y$ in y-direction (µm) |
|---|---|---|---|---|---|
| Shift Register 1 | 36 | 40 | 4 | 1 | 15 |
| Shift Register 2 | 39 | 40 | 1 | 1 | 15 |
| Shift Register 3 | 96 | 100 | 4 | 1 | 15 |
| Shift Register 4 | 36 | 40 | 4 | 1 | 15 |
| Shift Register 5 | 36 | 40 | 4 | 1 | 15 |
| Shift Register 6 | 36 | 40 | 4 | 1 | 15 |

TABLE II
FEATURES CIRCUITS WITH SLIT-TYPE MOATS

| Circuit FLXTRP2 | Moat length, $l$ (µm) | Moat pitch $p_x$ (µm) | Bridge between moats (µm) | Moat width, $w$ (µm) | Moat pitch $p_y$ (µm) | Superconducting layers below M4 ground plane |
|---|---|---|---|---|---|---|
| Shift Register 1 | 36 | 40 | 4 | 1 | 15 | M2 copy of M4 |
| Shift Register 2 | 39 | 40 | 1 | 1 | 15 | M2 copy of M4 + Fill array |
| Shift Register 3 | 96 | 100 | 4 | 1 | 15 | M2 copy of M4 + Fill array |
| Shift Register 4 | 36 | 40 | 4 | 1 | 15 | M1 copy of M4 + Fill array |
| Shift Register 5 | 36 | 40 | 4 | 1.5 | 15 | Only fill array |
| Shift Register 6 | 36 | 40 | 4 | 0.5 | 15 | Only fill array |

TABLE III
CIRCUITS WITH SQUARE, RECTANGULAR, AND SLIT MOATS

| Circuit FLXTRP4 | Moat size $a$ or $l$ (µm) | Moat size $b$ or $w$ (µm) | Moat pitch $p_x$ (µm) | Moat pitch $p_y$ (µm) | Superconducting layers below M4 ground plane |
|---|---|---|---|---|---|
| Shift Register 1 | 5 | 5 | 15 | 19 | Fill array |
| Shift Register 2 | 5 | 5 | 10 | 19 | Fill array |
| Shift Register 3 | 3 | 3 | 10 | 17 | Fill array |
| Shift Register 4 | 3 | 3 | 15 | 17 | Fill array |
| Shift Register 5 | 4 | 4 | 10 | 18 | Fill array |
| Shift Register 6 | 4 | 4 | 15 | 18 | Fill array |
| FLXTRP5 | | | | | |
| Shift Register 1 | 5 | 5 | 20 | 19 | Fill array |
| Shift Register 2 | 5 | 4 | 10 | 19 | Fill array |
| Shift Register 3 | 19.5 | 0.3 | 21.5 | 17 | Fill array |
| Shift Register 4 | 9 | 0.3 | 10 | 17 | Fill array |
| Shift Register 5 | 4 | 4 | 20 | 18 | Fill array |
| Shift Register 6 | 4 | 3 | 20 | 18 | Fill array |

TABLE IV
CIRCUITS WITH MULTIPLE GROUND PLANES AND VARIOUS FILL STRUCTURES BELOW M4 GROUND PLANE

| Circuit FLXTRP3 | Moat length, $l$ (µm) | Moat pitch $p_x$ (µm) | Bridge length (µm) | Moat width, $w$ (µm) | M4 and M7 width, $W$ (µm) | Superconducting layers below M4 ground plane |
|---|---|---|---|---|---|---|
| Register 1 | 36 | 40 | 4 | 1 | 14 | M1 copy of M4 + wire arrays on M0, M2, M3 |
| Register 2 | 39 | 40 | 1 | 1 | 14 | M1 copy of M4 + wire arrays on M0, M2, M3 |
| Register 3 | 96 | 100 | 4 | 1 | 14 | M1 copy of M4 + wire arrays on M0, M2, M3 |
| Register 4 | 36 | 40 | 4 | 1 | 14 | M0+M2 copy of M4 + wire arrays on M1, M3 |
| Register 5 | 36 | 40 | 4 | 1 | 14 | M3 copy of M4 + wire arrays on M0, M1, M2 |
| Register 6 | 36 | 40 | 4 | 1 | 14 | M1 copy of M4 + wire arrays on M0, M2, M3 |
| FLXTRP6 | | | | | | |
| Register 1 | 36 | 40 | 4 | 1 | 14 | M3 copy of M4 + array of wires on M0, M1, M2 |
| Register 2 | 39 | 40 | 1 | 1 | 14 | M3 copy of M4 + fill array |
| Register 3 | 96 | 100 | 4 | 1 | 14 | M3 copy of M4 + fill array + array of wires on M0, M1, M2 |
| Register 4 | 36 | 40 | 4 | 1 | 14 | M2 and M3 copies of M4 + array of wires on M0, M1 |
| Register 5 | 36 | 40 | 4 | 1 | 14 | Fill array |
| Register 6 | 36 | 40 | 4 | 0.3 | 14 | Fill array |

The circuit uses two ground planes, M4 and M7 in Fig. 4, connected by superconducting vias. Parameters of the moats in each of the six shift registers in the FLXTRP1 base circuit are given in Table I. The standard fill structure used consisted of a stacked 6 µm x 6 µm squares patterned in each Nb layer M0, M1, M2, and M3, and connected by vias; it was placed as an array with 10 µm x 10 µm pitch under the bottom ground plane M4 of each of the six shift registers. It is referred to hereafter as fill array. Four moat widths were investigated using the FLXTRP1 circuit and its modifications shown in Tables I and II: 0.3, 0.5, 1.0, and 1.5 µm.

*B. Circuits Using Square and Rectangular Moats*

By stretching the U-turn cell in the y-direction and changing the placement pitch of the rows to 17, 18, and 19 µm, we created a few circuits for testing ordered patterns of square and rectangular moats of various sizes $a \times b$ and pitch $p$ in the x-direction along the rows, placed between the adjacent rows of the shift registers as shown in Fig. 2a. The moat sizes and pitches are given in Table III. The moats were congruent in the M4 and M7 ground planes. For x-direction pitches non-commensurate with the 20 µm cell length, the relative positions of the moats and the cells change along the row.

*C. Shift Registers Using Multiple Ground Planes and Various Fill Structures Below M4 Ground Plane*

All estimates of the full flux expulsion field given in Sec. II are for the single superconducting film. The effect of using multiple, electrically connected, ground planes or the presence of patterned superconducting layers below or above the ground plane(s) on flux trapping is not known and has not been studied probably because of a limited number of layers in the fabrication technologies at the time. However, modern fabrication technologies have eight or more superconducting layers and their effects on flux trapping need to be evaluated and understood. Therefore, we have designed multiple versions of the FLUXTRP1 chip in which we used additional "dummy" ground plane(s) below the main M4 ground plane of the circuits. That is, a replica of the M4 was placed either on the M3 layer or on the M2, M1, or M0 layers, thus creating shift registers with three ground planes: two electrically connected, M4 and M7, and one disconnected. Then, we created versions with four ground planes, placing copies of the M4 on two other layers. All the designed combinations are given in Tables II and IV. On other layers below M4, not used as dummy ground planes, we placed arrays of parallel 1 μm or 2 μm wide wires imitating passive transmission lines frequently used for clock and data transmission below the ground plane. The wires in the adjacent layers were running perpendicular to each other. The wires did not form any closed loops and, hence, could not trap flux anywhere except for in the wires themselves. This fill structure is referred to as a wire array in Tables III and IV.

TABLE V
RATE OF CHANGE OF THRESHOLD CLOCK AMPLITUDES

| Critical currents of JJs in cell / Rate of change [2] | $I_{c1}$ | $I_{c2}$ | $I_{c3}$ | $I_{c4}$ |
|---|---|---|---|---|
| $\partial(PL)/\partial I_{ci}$ | −1.93 | 4.40 | 2.46 | 0 |
| $\partial(NL)/\partial I_{ci}$ | 0 | −0.97 | −1.14 | 5.37 |

The most complicated fill structure was a mixture of the wire fill and the standard array fill. In this case, the stacked 6 μm × 6 μm superconducting squares connected the wires running in one of the layers and created a pattern of superconducting loops below the ground plane.

*D. Circuit Testing and Extraction of Cell Operating Margins*

Measurements of the circuits were performed at a temperature of 4 K in a pulse tube cryocooler containing a single mu-metal shield. The ambient magnetic field within the shield was measured to be approximately 1.2 μT, using on-chip SQIF magnetometers. The setup allowed to test up to 28 5 mm × 5 mm chips wire-bonded onto a custom interposer PCB, and connected directly to room temperature connectors through 12 custom 50-pin Kapton flex cables. All measurements were performed using a 128-channel Octopux [53] at a frequency of approximately 10 kHz. All six shift registers on a single chip were measured in parallel, reducing the time required to characterize all cells on the chip by a factor of six.

The current bias for all DC-SFQ and SFQ-DC converters was supplied through a parallel biasing network. The operating range for the converters was 6.5 mA to 10.5 mA. For proceeding measurements, the current bias was set to 9.0 mA. The shift registers were operating correctly in the range of sinusoidal clock amplitude typically from ~0.5 mA to ~2.5 mA. That is, all cells within each shift register were found to be operational for clock amplitudes in this range that, hence, represents the circuit's global clock margins. This range is effectively a characterization of the worse-performing cell(s) within the shift register.

The operational range for clock amplitudes of individual cells within a shift register was measured by selectively modifying either the positive or negative amplitude of a particular half-period of the ac clock while the cell under test is in a logical "1" state, i.e., contains a single flux quantum as described in [2], [5]. Each cell can be characterized by four amplitudes: a) positive lower (PL) is a positive threshold amplitude correctly advancing logical state "1" from the first part of the cell, the loop $J1$-$L2$-$L3$-$J2$ in Fig. 4, to the second part of the cell into the second loop $J3$-$L5$-$L6$-$J4$; b) positive upper (PU), the highest amplitude for the correct operation; c) negative low (NL) – the smallest negative clock amplitude on the negative half period correctly advancing logical "1" from the second loop to the next cell in the register; d) negative upper (NU) - the largest negative amplitude for advancing "1" into the next cell in the register.

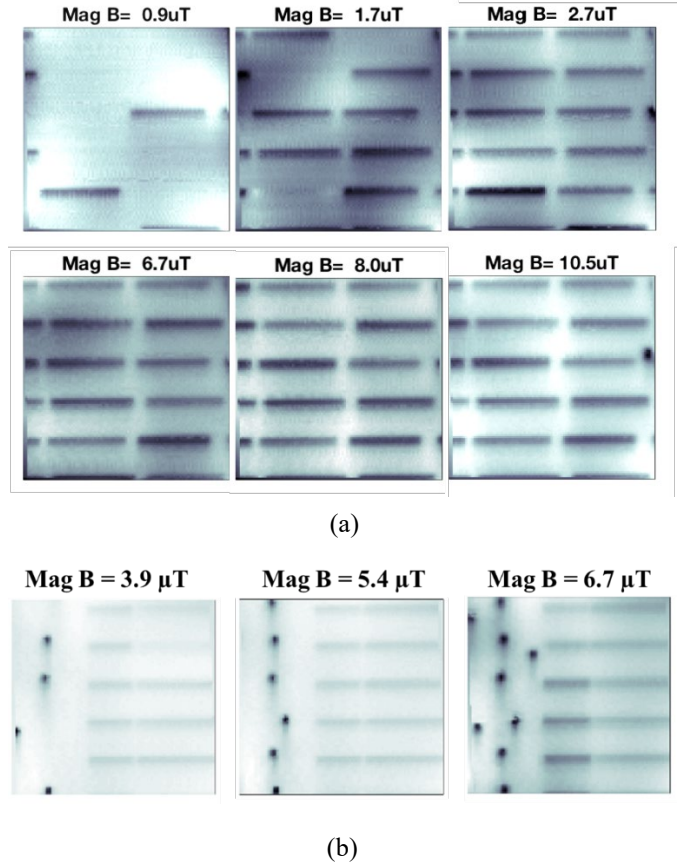

(a)

(b)

Fig. 7. a) Scanning SQUID microscope images of a 200-nm thick Nb ground plane with the same moats as in the FLXTRP1 circuit design: $l$ =36 μm with 4 μm bridges, $w$ =1 μm, and distance between the parallel rows of moats of $W$ =18 μm. Moats with trapped flux look dark, moats without flux are not visible. All moats in the field of view become filled with flux at a field of 2.7 μT. A black dot appearing near the moat corner in the image taken at 10.5 μT is an Abrikosov vortex trapped in the film because all moats have been fully filled and cannot sequester more flux. b) Magnetic flux images taken in the film area adjacent to the moats in (a) but having no moats of its own. Here vortices trapped in the film appear at a much lower field between 1.7 μT and 2.7 μT. Note that vortices tend to be located near corners of the moats.

Clock amplitude thresholds of all cells in the register can be measured using the procedure described in [2], [5]. In the simplest case, to probe the $k$-th cell, $k$ clock periods with the clock amplitude within the circuit global margins, the nominal amplitude, are applied to shift logic "1" into the cell. Then, a modified clock amplitude is applied, producing either a correct or incorrect shift of the data. It is checked by shifting the cell output to the end of the register, using the nominal clock amplitude. Timing errors in output data arrival with respect to the $N − k$ clock periods indicate incorrect operation at the location of the $k$-th cell; $N$ is the number of bits in the register

*E. Circuits for Testing Junction's $I_c$ Deviations*

The very same shift register circuit can be used to detect fabrication-induced changes in critical currents of Josephson junctions due to a direct relationship between the junction critical currents and the PL and NL threshold amplitudes [2]. From the simulations done in [2], it was established that variations of the critical current of junction $J2$, $I_{c2}$ have the largest impact on the lower positive threshold, about 4.4 μA change of the threshold per 1 μA change of the critical current. Similarly, the critical current of junction $J4$ has the largest impact on the negative lower threshold. The simulated rates of the PL and NL threshold changes with changes in the $I_c$ of each of the four JJs in the unit cell are given in Table V.

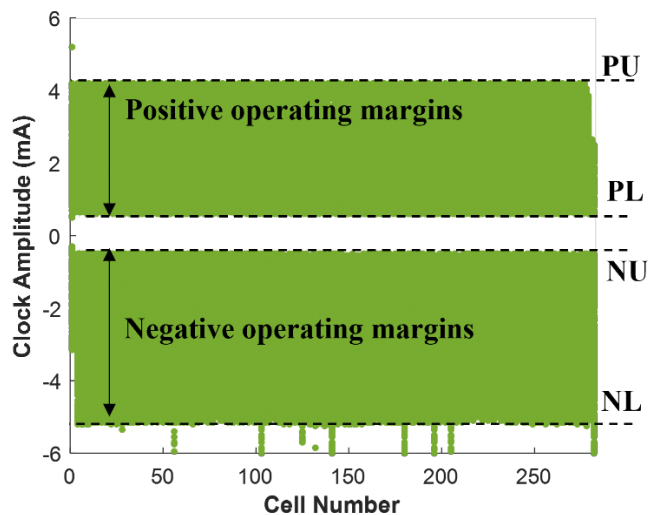

Fig. 8. Operating ranges of the positive and negative clock amplitudes (margins) of the first 282 individual cells in one of the shift registers in FLXTRP1 circuit. The positive lower (PL), positive upper (PU), negative lower (NL) and negative upper (NU) threshold amplitudes are defined in the text. The ac clock line is inductively coupled to the individual cells in the registers via transformers, $M$ shown in Fig. 2. The clock threshold values shown are the respective current amplitudes in the primary of the transformers. The first cell in the register has lower absolute values of the PL, NU and NL thresholds for correct operation than other cells due to an additional biasing from the Input DC-SFQ converter. At absolute values of clock amplitude higher than PU or NL a cell dose not operate correctly and advances the data father away along the register than is required, more than a half-cell. At absolute values of clock amplitude below PL or NU, a cell does not advance the data "1" along the register; see [2].

To use the shift registers as a reliable fabrication process diagnostic circuit and verify its high sensitivity to changes in the JJ critical currents, we designed a modification of the base circuit in which all junctions $J1$ in one particular row were replaced by junctions with a 10% higher critical current, i.e., 10% larger area, and in another row with 10% lower $I_{c1}$ of the $J1$. The same changes were introduced to other junctions $J2$, $J3$, and $J4$ using different rows in the six available resisters on the same chip. Unmodified rows served as a reference to establish the base PL and NL thresholds.

## IV. Test Results

### A. Magnetic Imaging of Flux in Moats

To estimate the flux explosion field for the ground plane configuration with the slit moats in Fig. 1, SQUID scanning microscopy images of the ground plane were taken after field cooling at different values of magnetic field, similarly to [37], [38], [30]. The typical images are shown in Fig. 7. The moats' configuration is the same as in the FLXTRP1 circuit and shown in Fig. 1, $W$ =18 μm. A full expulsion magnetic flux density $B_{exp}$ appears to be close to 10 μT. At $B_r$ =10.5 μT, an Abrikosov vortex becomes visible in the film near the right corner of one of the moats. Fig. 7b shows a magnetic flux image taken from the area adjacent to the patterned moats, farther away from the left side of the circuit. Note that in almost all cases Abrikosov vortices at $B_r \gtrsim B_{exp}$ get trapped near the corners of the moats, and not in the wider film area between the adjacent moats. This is likely because the Meissner screening currents circulating the flux-filled moats have the largest current density and gradient near the moat corners and, hence, the largest repulsive force on the vortex of the same polarity; see a sketch Fig. 1.

The observed $B_{exp}$ is consistent with (4) giving $B_K \approx$ 10.5 μT, whereas (2) and (3) give, respectively, $B_{c1} \approx$ 4.1 μT and $B_0 \approx$ 5.0 μT. In any case, the observed expulsion field for this moat configuration is much higher than the residual magnetic flux density $B_r$ in the test setups used for the measurements of the shift registers: $B_r \approx$ 0.2 μT and 1.2 μT in the LHe immersion probe and in the cryocooler, respectively. Hence, we can expect a very small moat filling factor of about 2% when using LHe immersion probe, about 12% filling when using the cryocooler-based setup, and no bad flux trapping of Abrikosov vortices inside the ground plane films based on these measurements and estimates. The presence of the second ground plane and of other superconducting structures below the bottom ground plane may change these expectations.

In a separate study, we investigated the dependence of the full expulsion field for 200-nm Nb film on the length of the moats at the fixed values of $w$=1 μm, $W$ = 14 μm, and pitch $p_x$= 40 μm by increasing the length of the bridge between the moats in Fig. 1. The results show a linear increase in the expulsion field from $B_{exp} \approx$1.5 μT at $l$ =10 μm with the slope $dB_{exp}/dl \approx$ 0.36 μT μm$^{-1}$ [52]. This indicates the advantage of increasing the length of the moats as much as allowed by the requirements of the circuit connectivity and routing in the $y$-direction, perpendicular to the moats.

### B. Margins of Individual Cells in the Base Circuit FLXTRP1

The FLXTRP1 chip was fabricated on multiple fabrication runs. Numerous copies of the chip were measured to extract the standard operating margins of the global clock, individual cells, and DC-SFQ and SFQ-DC converters. Out of 12 FLXTRP1 circuits, i.e., twelve 5 mm x 5 mm chips containing 72 shift registers with more than 1.3M JJs, all circuits were found to be fully operational.

An example of the individual cell margins for the first 282 cells in one of the shift registers is shown in Fig. 8. The operating range of most cells is approximately from 0.6 mA to 4.2 mA for the positive clock phase and approximately from 0.5 mA to 5.1 mA for the negative clock phase. These margins characterize the ac current amplitude in the primary of the transformer $M$ in Fig. 5 required to switch the designated junctions within the cell exactly one time to correctly propagate the data encoded by a single flux quantum presence (logical "1") or absence (logical "0") in the cell.

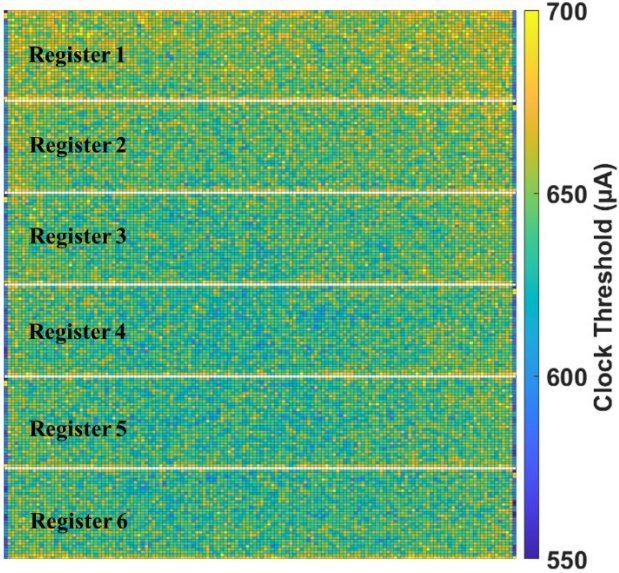

**Fig. 9.** Color ("heat") map of the positive lower (PL) threshold amplitude for the correct operation of all individual cells in all six registers in one of FLXTRP1 circuits. The PL thresholds for 6×4513=27078 cells have been measured using the method described in III.C and [2]. The ac clock line is inductively coupled to the individual cells in the registers via transformers, $M$ shown in Fig. 2. The clock threshold values shown are the positive half-period current amplitudes in the primary of the transformers. In registers 1, 3, and 5, the cell count starts from the top left corner and goes to the right, then down and left in a meandering fashion. Registers. 2, 4, 6 are mirror inverted with respect to the y-axis of the odd-numbered registers. Note that U-turn cells along the left and right edges of the registers have lower PL threshold clock amplitudes than cells in the rows by design.

It can be observed that the operating margins of the "special" cells described in Sec. III within the shift register are noticeably smaller than for the nominal cells. Due to the physical and operational differences, these 34 special cells were excluded from the statistical analysis of the individual cell in the shift registers. The distributions of margins with and without the special cells can be observed in Fig 11.

There are four main contributions to statistical variations of the threshold amplitudes: a) fabrication-induced differences in the critical currents of JJs and inductors in different cells; b) flux trapping in some of the moats; c) thermal noise; d) instrumental (test setup) noise [2]. Since they are independent, the total variance of the threshold clock amplitude in a register,

$$\sigma_{tot}^2 = \sigma_{fab}^2 + \sigma_{flux}^2 + \sigma_{th}^2 + \sigma_{exp}^2, \qquad (9)$$

is the sum of variances of these random contributions. Repeatable measurements of the threshold amplitudes of the same cell, using 1 μA step size, allowed us to estimate $\sigma_{th}^2 + \sigma_{exp}^2$ because in this case $\sigma_{fab}^2 = 0$; the flux distribution in the moats also does not change and, hence, $\sigma_{flux}^2 = 0$. These measurements gave $\sigma_{th}^2 + \sigma_{exp}^2 \approx 9$ μA$^2$.

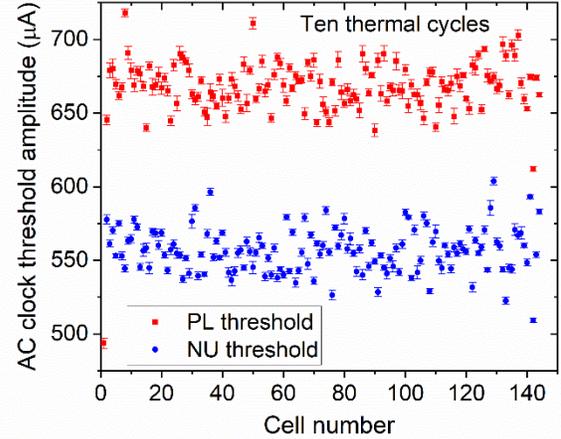

**Fig. 10.** Variations of the measured positive low (PL) and negative upper (NU) clock amplitude thresholds over ten thermal cycles for the first 145 cells in a single FLXTRP1 shift register, register #2. The data points show the mean values, $\mu_i$ and the error bars correspond to one standard deviation $\pm\sigma_{flux}$ of the cell threshold variations in ten thermal cycles. For the PL thresholds, the largest variation observed was $\sigma_{flux} \approx 6$ μA, $\sigma_{flux}/\mu \approx 0.9\%$. For the NU thresholds, the largest variation was $\sigma_{flux} \approx 4$ μA, $\sigma_{flux}/\mu \approx 0.7\%$.

We estimated $\sigma_{flux}^2$ by performing thermal cycling after measuring the cells margins and then repeating the margins measurements. In this case, each cooldown produces a new flux distribution, so repeating this cycle multiple times gives a measure of $\sigma_{flux}^2$. From these measurements, the maximum value of the moat flux-induced PL threshold variance was $\sigma_{flux}^2 \approx 36$ μA$^2$. The typical results are shown in Fig. 10 for the first row of cells in a register for better visibility. It is clear that changes in the clock threshold values for any particular cell as a result of thermal cycling are much smaller than the difference in the clock threshold values for different cells. This indicates the prevailing contribution of $\sigma_{fab}^2$, the fabrication-induced differences between the cells in the register.

The distributions of the PL threshold amplitudes for all the 4481cells in the rows (4513 bits minus 32 special cells) of one of the six registers in the FLXTRP1 circuit are given in Fig. 11. To save testing time, the measurements were done using a 10 μA step size for the clock current increment instead of 1 μA. This increases $\sigma_{exp}$ to 10 μA. We used a simple Gaussian fit to characterize these distributions, although they appear to be slightly skewed to the right, i.e., having more cells with higher threshold amplitudes than with lower values than the mean. From the Gaussian fit, $\sigma_{tot} = 16$ μA and $\sigma_{tot}/\mu = 2.5\%$. These values are very close to the values obtained in [2], $\sigma_{tot} = 14.7$ μA and $\sigma_{tot}/\mu = 3.3\%$, for the PL clock amplitude thresholds for the cells in a similar ac-powered register with 16k cells fabricated nine years ago in the nominally the same process.

Deducting $\sigma_{th}^2 + \sigma_{exp}^2 = 100$ μA$^2$ set by the measurements step size and $\sigma_{flux}^2 = 36$ μA$^2$, we get $\sigma_{fab} = 11$ μA and $\sigma_{fab}/\mu = 1.7\%$. The root-mean-square (rms) variation of the PL threshold, $rms(PL)$ is related to the root-mean-square variation of the critical currents of JJs in the cells as $rms(I_c) \approx rms(PL)/5.4$; see [2] and Table V. Then, the fabrication-induced variation of the JJ critical currents in the studied register is $\sigma_{JJ} = \sigma_{fab}/5.4 = 2$ μA or 1.6% of the nominal $I_c = 125$ μA of the JJs used in the registers. The former value is extremely close to the normalized standard deviation $\sigma_{JJ}/\mu_{Ic}$ of the critical currents extracted from the $I_c$ and normal resistance measurements of the individual JJs of the same size [48], [49], [55], although statistics in such measurements is limited to only a few hundred JJs whereas the data in Fig. 11 are based on almost 18k JJs.

TABLE VI
EFFICIENCY OF SLIT-TYPE MOATS IN PROTECTING SHIFT
REGISTERS AGAINST FLUX TRAPPING OUTSIDE THE MOATS

| Length, $l$ (μm) | Pitch $p_x$ (μm) | Width, $w$ (μm) | Pitch $p_y$ (μm) | Total number of cooldowns | Fully operational after cooldown | Fully operational (%) | Probability of bad flux trapping (%) |
|---|---|---|---|---|---|---|---|
| 36 | 40 | 1 | 15 | 486 | 476 | 98 | 2 |
| 39 | 40 | 1 | 15 | 117 | 115 | 98 | 2 |
| 96 | 100 | 1 | 15 | 122 | 116 | 95 | 5 |
| 36 | 40 | 1.5 | 15 | 68 | 64 | 94 | 6 |
| 36 | 40 | 0.5 | 15 | 65 | 63 | 97 | 3 |
| 36 | 40 | 0.3 | 15 | 20 | 20 | 100 | 0 |
| 9 | 10 | 0.3 | 17 | 45 | 45 | 100 | 0 |
| 19.5 | 20.5 | 0.3 | 17 | 45 | 45 | 100 | 0 |
| | | | Total: | 968 | 944 | Mean: 97.5 | 2.5 |

TABLE VII
EFFICIENCY OF SQUARE MOATS IN PROTECTING SHIFT
REGISTERS AGAINST FLUX TRAPPING OUTSIDE THE MOATS

| Length, $a$ (μm) | Width, $b$ (μm) | Pitch, $p_x$ (μm) | Pitch $p_y$ (μm) | Total number of cooldowns | Fully operational after cooldown | Fully operational (%) | Moat number density, $n_{moat}$ (cm$^{-2}$) |
|---|---|---|---|---|---|---|---|
| 3 | 3 | 10 | 17 | 45 | 45 | 100 | 588,235 |
| 3 | 3 | 15 | 17 | 45 | 45 | 100 | 392,157 |
| 4 | 3 | 20 | 18 | 45 | 45 | 100 | 277,777 |
| 4 | 4 | 10 | 18 | 45 | 45 | 100 | 555,555 |
| 4 | 4 | 15 | 18 | 45 | 45 | 100 | 370,370 |
| 5 | 5 | 10 | 19 | 45 | 45 | 100 | 526,316 |
| 5 | 5 | 15 | 19 | 45 | 45 | 100 | 350,877 |

In the estimates above, we neglected the effect of variations of the cell inductors on the thresholds. In simulations, this effect is much smaller than the influence of the JJ variations; moreover, inductance variations themselves are very small, less than 1% for 0.4-μm-wide cell inductors [45], [46], [54].

So, we established that the observed variations in the characteristics of the individual cells in the registers are dominated by the fabrication-induced variations of the critical currents of Josephson junctions. The effect of good flux trapping in the moats, the flux sequestration, on the individual cell margins was found to be very small, about 1%, as was expected from the circuit design and in agreement with numerical simulations of the cell inductors coupling to fluxons in the moats [42].

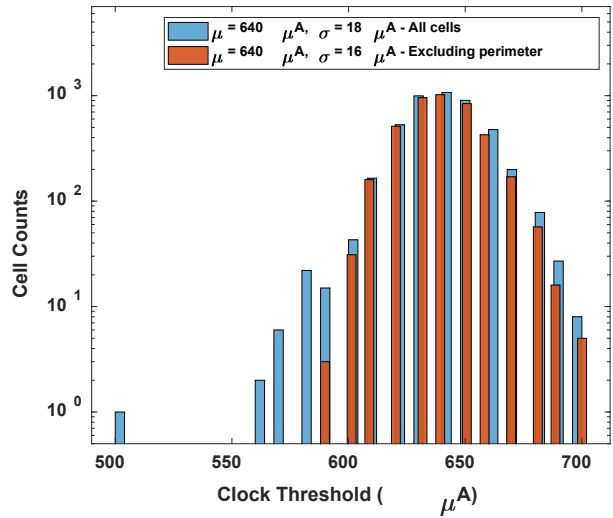

Fig. 11. Distribution of the positive clock threshold for the individual cells within a single shift register. The histogram in blue contains all 4513 cells within the shift register, while the histogram in red excludes all turn cells and I/O cells, and represents 4481 cells in the rows.

*C. Moat Geometry Effect on Flux Sequestration Efficiency*

We investigated the efficiency of different moat geometries in protecting against flux trapping inside the circuits by using shift registers with moats described in Tables I - III and measuring the functionality of the circuits over multiple thermal cycles. In this characterization, a shift register was deemed fully functional on a particular cooldown if its global operating margins were within ±10% of the nominal value of the benchmark FLXTRP1 shift register. This criterion was chosen because ±10% shift in the margins is much larger than any shift expected from flux sequestered in the moats. A larger shift, definitely indicating flux trapping outside of the moats, was categorized as bad flux trapping. Although a 10% change in the margins does not render the studied registers nonoperational, it can be detrimental for other types of circuits with much smaller operating margins. All cooldowns were done in the cryocooler-based setup in the nominally the same residual magnetic field of about 1.2 μT and with the same cooling rate 0.5 K/min.

The results of this investigation are listed in Table VI for the slit-type moat of different length and width. Overall, varying dimensions of the moats within the selected parameter range had a minimal impact on the functionality of the measured shift registers. The average probability of a noticeable flux trapping outside of the slit-type moats was found to be 2.5%. Small differences in flux trapping probability between moats of different sizes in Table VI could be explained by differences in the statistics – the number of cooldowns. Indeed, at a 2.5% average flux trapping probability, observing a single bad flux trapping event would require at least 40 cooldowns on average, more than was used in some cases. We have not observed any detrimental flux trapping in FLXTRP1 circuits designed with long slit-type moats in numerous field cooling cycles done with different cooling rates from 0.1 K/min to 5 K/min and on a large

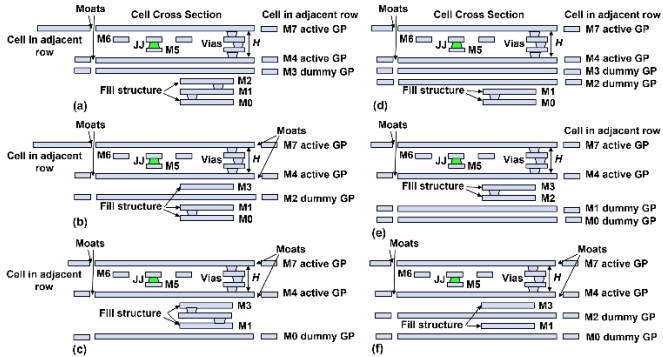

**Fig. 12.** Schematic cross sections of the shift register cells with JJs and inductors located between electrically connected active ground planes M4 and M7 with moats described in Table I, and various additional "dummy" ground planes on layers from M0 to M3 below the M4 layer: (a)-(c) single dummy GP on layer M3 or M2, or M0; (d)-(e) – dummy GPs on two adjacent layers below the M4; (f) – dummy GPs on two nonadjacent layers below the M4. Some of the studied combinations of the dummy GPs are not shown as well as dummy GPs on three and four layers; see Table VIII. Fill structure patterns are 6 μm x 6 μm on 10 μm pitch. Dummy GPs are 14 μm wide strips unconnected to the adjacent rows and other layers.

number of circuits fabricated in different fab runs.

The excellent performance of the long slit-type moats in preventing bad flux trapping agrees with the scanning SQUID images in Fig. 7, where the full expulsion field of 40 μm long moats at $W = 18$ μm was found to be approximately 10 μT, a factor of 10x larger than the ambient magnetic field in the cryocooler-based test setup used.

*The most important finding is that the width of the moats, w can be reduced to the minimum process-allowed width*, e.g., 0.3 μm without affecting flux-trapping performance of the moats. This significantly reduces the negative impact of the moats area on the circuit density. In the future, it should be possible to use much thinner cuts in the ground planes for the flux sequestration because even an infinitely thin cut is sufficient to interrupt screening currents circulating the vortices and create an attraction force.

Quite unexpectedly, square and rectangular moats with sizes from 3 μm x 3 μm to 5 μm x 5 μm and pitches from 10 μm to 20 μm *turned out to be as efficient in protecting the shift registers against bad flux trapping as the slit-type moats*, based on 45 cooldowns; see Table VII and Fig. 2a. This is an important finding because square-shaped moats are convenient for logic cell tiling and routing in large-scale circuits.

At the $B_r = 1.2$ μT in the test setup, the expected number density of vortices needed to be sequestered or expelled is $n_v \approx 5.8 \cdot 10^3$ cm$^{-2}$, which is at least five times less than the number density of the square moats $n_{moat}$ used in the registers; see Table VII. Therefore, (5) is satisfied. For the square moats, (6) becomes $n_v a p_y \leq 1$. The latter is less than 0.055 for the cases studied. This means that the circuit designs provided many more moats than is sufficient to capture no more than one vortex per moat.

The studied square and rectangular moats occupy from 3.3% to 13.2% of the shift register area. If the former number is acceptable the latter is too large a waste of the circuit area. For comparison, slit-type moats with $w = 0.3$ μm occupy less than 1.8% of the total circuit area, and their area can be reduced further.

*E. Effect of Multiple Ground Planes on Flux Trapping*

All combinations of the ground planes (GPs) used in this study are given in Tables II and IV. All the circuits have two active ground planes, M4 and M7 connected by superconducting vias, below and above the JJs and wires forming stripline inductors with these GPs. These GPs are necessary and cannot be avoided. 'Dummy' GPs were added to study the effect of multiple GPs on flux trapping outside the moats in the M4 and M7. Schematic cross sections of the cells using a few of the possible combinations of 'dummy' ground planes studied are given in Fig. 12. Dummy ground planes present a copy of the M4 active ground plane but without bridges between the moats. They are stripes of Nb film with width $W = 14$ μm and 200 nm thickness, placed under each row of cells on the corresponding layer in one or several shift registers of the FLXTRP2, FLXTRP3, and FLXTRP6 chips described in Tables II and IV. The dummy GPs are completely passive and electrically floating: they have no electrical connections to any circuitry and no connections between themselves and to the active ground planes. The summary of the results on bad flux trapping rendering shift registers containing multiple GPs nonoperational is given in Table VIII.

The first striking result of these measurements is a dramatic

TABLE VIII
EFFECT OF MULTIPLE GROUND PLANES ON PROBABILITY OF BAD FLUX TRAPPING

| Active Ground Planes | Dummy Ground Plane(s) | Total number of cooldowns | Fully operational after cooldown | Fully operational (%) | Distances between ground planes, $H$ (nm) |
|---|---|---|---|---|---|
| M4 and M7 | none | 421 | 415 | 99 | $H_{M4-M7} = 1015$ |
| M4 and M7 | M0 | 20 | 20 | 100 | $H_{M0-M4} = 1400$ |
| M4 and M7 | M1 | 526 | 503 | 96 | $H_{M1-M4} = 1000$ |
| M4 and M7 | M2 | 215 | 107 | 50 | $H_{M2-M4} = 600$ |
| M4 and M7 | M3 | 208 | 4 | 1.9 | $H_{M3-M4} = 200$ |
| M4 and M7 | M1+M0 | 20 | 20 | 100 | $H_{M1-M4} = 1000$ $H_{M0-M1} = 200$ |
| M4 and M7 | M2+M0 | 148 | 20 | 14 | $H_{M2-M4} = 600$ $H_{M0-M2} = 600$ |
| M4 and M7 | M2+M1 | 20 | 0 | 0 | $H_{M2-M4} = 600$ $H_{M1-M2} = 200$ |
| M4 and M4 | M3+M1 | 20 | 0 | 0 | $H_{M3-M4} = 200$ $H_{M1-M3} = 600$ |
| M4 and M4 | M3+M2 | 40 | 0 | 0 | $H_{M3-M4} = 200$ $H_{M2-M3} = 200$ |
| M4 and M4 | M2+M1+M0 | 20 | 0 | 0 | $H_{M2-M1} = H_{M1-M0} = 200$; $H_{M2-M4} = 600$ |
| M4 and M7 | M3+M2+M1 | 20 | 0 | 0 | $H_{M3-M4} = H_{M3-M2} = H_{M2-M1} = 200$ |

TABLE IX
MEASURED AND CALCULATED CHANGES OF THRESHOLD CLOCK AMPLITUDE INDUCED BY MODIFICATIONS OF JJS

| Junctions $I_c$ modification | $\delta I_{c1}/I_{c1} = 0.1$ | $\delta I_{c2}/I_{c2} = 0.1$ | $\delta I_{c3}/I_{c3} = 0.1$ | $\delta I_{c4}/I_{c4} = 0.1$ | $\delta I_{c1}/I_{c1} = -0.1$ | $\delta I_{c2}/I_{c2} = -0.1$ | $\delta I_{c3}/I_{c3} = -0.1$ | $\delta I_{c4}/I_{c4} = -0.1$ |
|---|---|---|---|---|---|---|---|---|
| Measured change in PL threshold (%) | −4.2 | 12.5 | 3.2 | −1.9 | 4.5 | −12.3 | −4.8 | 0.6 |
| Calculated change in PL threshold (%) | −3.8 | 8.7 | 4.8 | 0 | 3.8 | −8.7 | −4.8 | 0 |

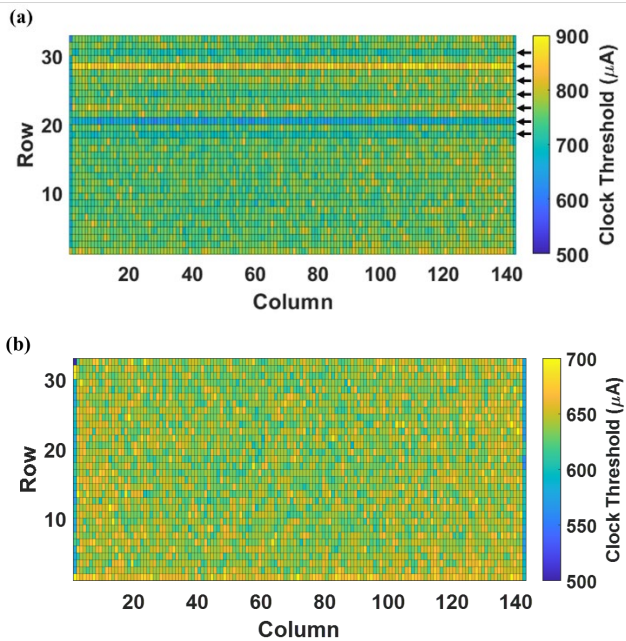

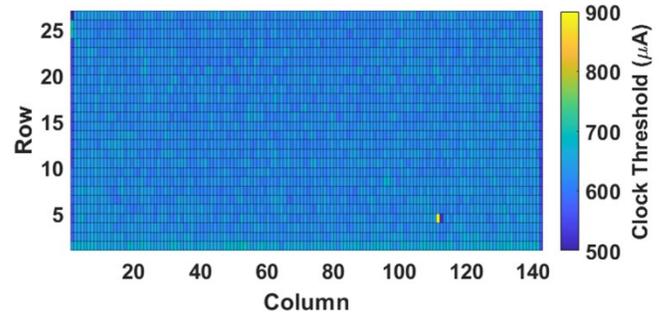

Fig. 14. Color map of clock threshold amplitudes PL for cells in the shift registers. An individual cell with a significantly shifted clock threshold can be observed near column 112, row 4. A cell adjacent to the impacted cell also has a slightly shifted clock threshold.

Fig. 13. Color map of threshold clock amplitudes PL for cells in the shift registers. (a) $J$1 areas in the entire row of cells shown by arrows, from top to bottom, was increased by 15%, increased by 10%, reduced by 15% (intense blue row # 21), and reduced by 10%. The remaining cells within the shift register are unmodified. (b) The base shift register with unmodified cells.

$\frac{1}{N}\Sigma_i((PL)_i/\mu_{PL} - 1))$ for each modified row, where summation is over all cells in the modified row and $\mu_{PL}$ is the mean PL threshold amplitude of the unmodified cells. The results are shown in Table IX along with the calculated relative increase in flux trapping probability outside of the moats with decreasing distance between the bottom active ground plane M4 and the third (dummy) ground plane below it. At the smallest distance $H_{M3-M4} = 200$ nm allowed in the fabrication process, using the M3 GP, the shift registers became completely nonfunctional as a result of flux trapping. However, remote ground planes distanced by more than 1000 nm, like M1 or M0, had no detectable effect on the circuit operation and flux trapping.

The second striking observation is a strong enhancement of detrimental flux trapping by pairs of dummy ground planes, especially closely spaced. For instance, any GP pair including the M3, or an M2+M1 pair, made the shift registers trap flux in 100% of cooldowns, rendering the circuits completely nonoperational. Only an M1+M0 pair of dummy GPs, separated from the M4 GP by 1000 nm, had no effect on flux trapping probability.

*F. Circuits with Intentionally Altered Josephson Junctions*

In order to confirm the high sensitivity of the clock threshold amplitudes to changes of the critical currents of individual junctions, e.g., $I_{c1}$ in Table V, we measured the individual cell margins of the shift registers with intentionally altered JJs described in III.E. The typical map of the PL clock amplitude thresholds for a register with modified junctions $J$1 in several rows of the register is shown in Fig. 13a whereas Fig.13b shows the map for the regular register with unmodified JJs for comparison. The rows of cells with modified JJs are easily detectable. The same measurements were repeated on the registers with rows of cells with modified junctions $J$2, $J$3, and $J$4 in Fig. 5.

Based on the measurements, the average relative change in the PL clock threshold amplitudes was defined as $\delta(PL)/\mu_{PL} =$

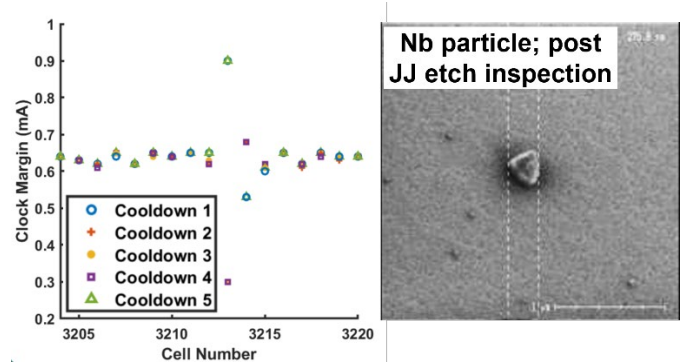

Fig. 15. Left panel: Measurement of clock threshold amplitudes over multiple thermal cycles for the shift register containing an anomalous cell. The clock PL threshold of the anomalous cell and, to a lesser extent, the adjacent to it cell vary widely between consecutive cooldowns. Right panel: SEM image of the typical defects found in the area of the shift register chip at defect inspection during the wafer fabrication.

changes in the threshold given by $\delta(PL)/\mu_{PL} = \left(\frac{\partial PL}{\partial I_{ci}}\right)\frac{\delta I_{ci}}{I_{ci}}$, where $\frac{\delta I_{ci}}{I_{ci}}$ is the relative change in critical current of the modified JJs and $\left(\frac{\partial PL}{\partial I_{ci}}\right)$ the impact coefficients from [2] and Table V. Agreement between the measured and calculated changes in PL thresholds is quite good. The same analysis for the negative thresholds NL showed similarly good agreement between the measurements and the simulations; it is not presented here to save space.

*G. Detecting Fabrication Defects*

Physical defects induced in circuit fabrication can impair the circuit functionality or become flux trapping sites. In the SFQ5ee process, wafers are optically inspected for defects larger in size than about 0.5 μm by a KLA-Tencor Altair defect inspection tool after each etching and chemical-mechanical planarization (CMP) fabrication steps. However, not all observed defects impact circuit performance, and small

<0.5 μm fabrication defects are usually not detected and analyzed.

Defects in the shift register circuits can potentially impact the individual cell margins by changing the cell inductances or junction critical currents. These defects will not change at thermal cycling. Defects on the surface of the ground planes, e.g., Nb particles and pinholes, can also serve as preferential flux pinning sites within the circuit. The impact of such defects may change from cooldown to cooldown. Both types of defects can potentially be detected through measurement of the individual cell margins.

Out of 168 shift registers tested with more than 3M JJs, we have found two registers demonstrating a behavior that can be attributed to the presence of fabrication defect(s) in one of the cells. One of them is shown in Figs. 14 and 15. Fig. 14 shows the positive lower thresholds for the cells in the register, demonstrating one outlier cell located in the fourth row in column 112. Its PL threshold is shifted significantly with respect to the rest of the cells. On repeated thermal cycles, the PL clock threshold (positive lower operating margin) of the cell varied significantly as shown in the left panel of Fig. 15, indicating flux trapping within the cell. Inspection of SEM images taken during the defect inspection after JJ top electrode etching step revealed niobium micro-masking defects, shown in the right panel of Fig. 15, located on this chip. Probably, regions with increased Nb thickness serve as pinning sites for vortices near JJs. Unfortunately, we were not able to confirm the presence of these defects in the affected cell by the chip inspection after the margins measurements because the cells are covered by the top Nb ground plane M7 and an $SiO_2$ passivation layer making the JJs and their surroundings invisible in optical and electron microscopes.

*H. Effect of the Ground Planes Track Width on Flux Trapping*

A modification of the FLXTRP1 circuit was created with registers having the moat pitch in $y$-direction of 15, 30, 45, and 60 μm, making the effective width $W$ of the ground plane tracks between the moats in Fig. 1 of 14, 29, 44, and 59 μm, respectively. No changes in the flux trapping probability were observed in the registers with $W \leq 44$ μm; and an increase in the number of unsuccessful cooldowns was observed in the register with $W = 59$ μm. These results are consistent with the vortex field expulsion (4) at the measured $B_r \approx 1.2$ μT in the test setup.

*I. Effect of Noncongruent Moats on Flux Trapping*

Two shift registers were designed with slit-type moats on the y-direction pitch $p_y = 30$ μm but with the y-offset between the moats in the bottom and top ground planes of 15 μm, i.e., the moats were not congruent but shifted significantly with respect to each other. These shift registers were found to trap flux in 100% of cooldowns, rendering them nonoperational.

V. DISCUSSION OF THE RESULTS

*A. Effect of Multiple Ground Planes on Flux Trapping*

In general, a superconducting GP in integrated circuits is needed to provide a low inductance path for the return currents in the circuit inductors and Josephson junctions, and to shield the circuits from the external magnetic field. However, a single GP does not screen the parallel magnetic field. Also, with a single GP, the circuit inductors are of a microstrip type, i.e., a wire above or below the GP, and have a long-range interaction between them − their mutual inductance decreases only as a second power of the distance between them [46]. Therefore, building complex, dense superconductor integrated circuits with a single GP is not possible, and two connected GPs shielding the circuitry from below and above are the most often used.

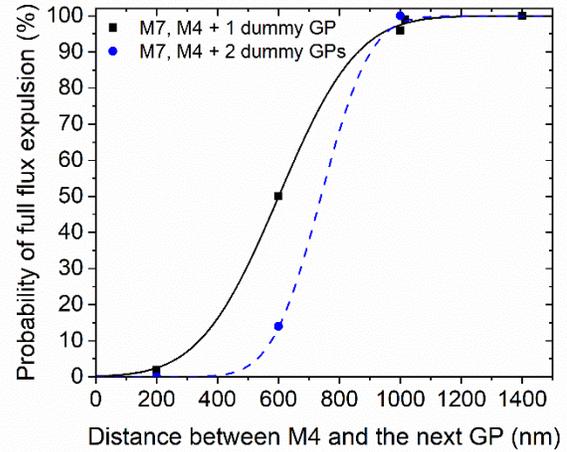

Fig. 16. Probability of the full flux expulsion from the shift register circuits as a function of the distance between the bottom ground plane M4 and the closest single dummy ground plane below it (■) and the pair of dummy ground planes (●). The width of the dummy GPs is 14 μm, the same as the distance $W$ between the slit-type moats in the M4 and M7 GPs. Curves connecting data points are to guide the eye. It is assumed that the total flux expulsion from the M4 and M7 active ground planes results in fully operational circuits while flux trapping outside the moats leads to a detrimental loss of the circuit operability; see the data in Table VIII.

Two GPs connected by superconducting vias can screen magnetic fields parallel to the GPs by superconducting currents circulating in loops between the GPs. Unfortunately, the same loops can trap magnetic flux parallel to the GPs upon cooling down in a residual magnetic field.

In very many cases, the third and fourth GPs may be required below or above the two mentioned GPs, e.g., to isolate two data routing tracks − passive transmission lines (PTLs) − going in mutually perpendicular directions, and/or to shield a dc or ac power distribution network − power bus − from the circuitry.

Without all these GPs, complex integrated circuits cannot be built. Therefore, understanding how these GPs affect flux expulsion and flux trapping is of a paramount importance.

The estimates of the flux expulsion field in II.A are applicable to a single layer of superconducting film stripes in a perpendicular magnetic field. Our results clearly indicate that the flux expulsion field and flux trapping probability change if there are multiple superconducting layers, and depend on the number of layers and distance between them, as shown in Fig. 16 and Table VIII.

Our results indicate that there is electromagnetic interaction between the superconducting layers. If the distance between the adjacent layers is large enough, $t \gg t_c$, they apparently behave as independent layers expelling fluxons to the moats independently. Presumably, the interaction between fluxons on different layers is weak in this case. However, at small distances between the adjacent layers and near $T_c$ where $\lambda(T) \gg t$, the two superconducting layers behave as one with an effective Pearl penetration depth $\Lambda/2$. The second film provides an additional screening of the stray fields produced by fluxons on the first film, reducing their interaction with the moats, and vice versa. Placing the third film at a distance $t \ll \lambda(T)$ adds more screening and reduces the penetration depth to $\Lambda/3$, and so on.

Fluxons of the same polarity on adjacent layers attract each other trying to align vertically to reduce the length of the field lines and electromagnetic energy in the dielectric between the layers. In this case, expulsion of flux into the moats requires synchronized movement of pancake vortices on different layers in the stack against the pinning forces in each of the layers.

Attraction of a vortex in a film to a moat or the film edge is mainly due to the currents near the metal/air and metal/substrate interfaces where the vortex field expands and the circulating currents decay as $r^{-2}$ [56]. For closely spaced layers in a multilayer, the vortex expansion occurs mostly near the very top and the very bottom interfaces. Hence, the attraction force to a moat does not increase significantly with increasing the number of layers, whereas the vortex pining force grows proportionally to the number of layers, i.e., the total thickness of the superconducting layers. Therefore, it should be easier to pin the vortex in a multilayer than in a single layer, i.e., the vortex freezing temperature of the multilayer should be higher than that of the single layer.

The flux expulsion becomes even more problematic if the moats in two superconducting layers are noncongruent and significantly offset with respect to each other. Flux expulsion in this case requires stretching the field lines between the aligned pancake vortices on the adjacent layers and increasing the total electromagnetic energy.

The described three-dimensional effects in the behavior of a stack of thin films are in many respects analogous to the flux coupling discovered in the 1960s in the so-called flux transformers, systems of two films with vortices [57], [58], and extensively studied theoretically since then [59], [60], [61], [62], especially in relation to layered high-$T_c$ superconductors; see [63] and references therein. Unfortunately, the theories of magnetic coupling and interlayer vortex interactions were developed for uniform multilayers consisting of infinite superconducting films and cannot be applied to a stack of strips or patterned films with moats. An attempt to develop a theory of Pearl's vortex interaction with a film edge was not very successful [24]. Magnetic field and current distribution in superconducting strips with slits in a perpendicular magnetic field was calculated in [64], providing an explanation for the increase of critical currents and magnetic hysteresis observed in films with slits and caused by enhancement of the surface barrier effects; see [64] and reference therein. Development of numerical methods for analyzing flux expulsion from stacks of realistic ground planes with various moat shapes is required.

*B. Effects of the Differences in $T_c$ and Film Nonuniformities*

The existing models of the flux expulsion from a single film and of magnetic interactions between different superconducting layers in a stack assume that all the films are uniform, have the same $T_c$, and ignore all cooling dynamic effects. However, in the simplest case of two films having different critical temperatures $T_{c2} > T_{c1}$ and different vortex freezing temperatures, $T_{f1}$ and $T_{f2}$, there is no interaction between the films in the temperature range $T_{c2} \geq T \geq T_{c1}$. If the difference in $T_c$s is sufficiently large, the film with the higher $T_c$ may fully expel flux before the film #2 becomes superconducting. Let us assume however that, because of the cooling dynamics, the full expulsion has not happened while the temperature of the lower-$T_c$ film reached $T_{c1}$. Then, the superconducting transition of the film #1 occurs in a nonuniform magnetic field created by the flux distribution in the film #2, imposing a different flux distribution than the one which would have happened if the film #1 transitioned superconducting on its own in the uniform field. Now, the flux needs to be expelled from two interacting films before the temperature of the film #2 becomes lower than the highest of the two vortex freezing temperatures. Even more complicated scenarios may take place in a stack of many films with different critical and vortex freezing temperatures, different film nonuniformities and vortex pinning strengths, and different widths of the superconducting transitions.

In reality, $T_c$ of Nb layers depends on the fabrication process sequence – the bottom layer, deposited and patterned first has the lowest $T_c$ due to the largest time it spends at elevated temperatures imbedded into the $SiO_2$ interlayer dielectric during processing the layers above it. As a results, the typical sequence of critical temperatures of the layers in the SFQ5ee process is $T_{c,M7} > T_{c,M6} > T_{c,M5} > T_{c,M4} > T_{c,M3} > T_{c,M2} > T_{c,M1} > T_{c,M0}$, and the difference $T_{c,M7} - T_{c,M0}$ can reach about 0.3 K. Therefore, ground planes M7 and M4, or M4 and M0, which have noticeably different critical temperatures, most likely expel magnetic flux independently of each other. However, any pair of the adjacent layers in the stack, e.g., M4 and M3, or M3 and M2, etc., most likely expel and pin flux collectively as a result of a smaller difference in $T_c$s and geometrical proximity, leading to a stronger flux trapping inside the films.

VI. CONCLUSION

We have demonstrated that superconductor very large-scale integrated circuits using two active Nb ground planes with 200 nm thickness and more than 1 μm distance between them can be fully protected against flux trapping using very narrow slit-type, congruent, moats occupying less than 2% of the circuit area and having negligible impact on the scale of integration.

We have found a strong enhancement of the detrimental flux trapping in the circuits using additional ground planes if spacing between any pair of the ground planes is less than about 0.6 μm. We attribute this to magnetic interaction between vortices on the layers having close critical temperatures and to an enhancement of flux pinning in a system of layers in comparison to a single layer.

The presence of noncongruent moats or a mix of congruent and noncongruent moats in the ground planes lead to a strongly enhanced flux trapping in the circuits.

The width of the ground planes tracks, i.e., the distance between the rows of parallel slit-type moats can be increased up to a certain level $W_c$, about 60 μm in our circuits, without noticeable enhancement of the flux trapping in the circuits. This level depends on the residual magnetic field, configuration of the moats, and their number density.

The presented results show that shift register cell margin measurements [2] can be used to efficiently characterize large quantities of JJs in integrated circuits and circumvents common challenges faced when attempting to measure large quantities of JJs at cryogenic temperatures using four-wire measurements of individual JJs and JJ arrays.

Some of the fabrication defects, which were detected by the optical defect inspection tool during the wafer fabrication and classified as nonkiller defects, do not make the circuit completely nonoperational but can cause significant flux trapping. They can be detrimental for the circuit operating margins and may require multiple thermal cycling to obtain operability. Combining our findings from the individual cell margin measurements in shift registers with the results of the defect inspection should allow as to develop better defect classification models and better understanding of defects causing flux trapping in superconductor integrated circuits.

ACKNOWLEDGMENT

The authors would like to thank Vladimir Bolkhovsky, Ravi Rastogi, and David Kim for their part in fabricating the circuits, and to Scott Meninger for the discussions of moat flux to inductors coupling and providing results of his simulations [42] prior to their publication. We are very grateful to John R. Kirtley and Kathryn Ann Moler for providing us with the scanning SQUID microscope images of Nb ground planes with patterned moats.